\documentclass[preprint]{aastex62}
\usepackage{graphicx}
\usepackage{rotating}
\usepackage{newunicodechar}

\newcommand{\kms}[1]{$#1$~km~s$^{-1}$}

\newunicodechar{Т}{T}
\received{August 29, 2020}
\revised{October, 14, 2020}
\accepted{October, 15, 2020}
\submitjournal{ApJ}

\shorttitle{Light Bridge Jets}
\shortauthors{Lim et al.}

\begin{document}

\title{Detection of Opposite Magnetic Polarity in a Light Bridge: Its Emergence and Cancellation in association with LB Fan-shaped Jets}

\email{eklim@kasi.re.kr}

\author[0000-0002-7358-9827]{Eun-Kyung Lim}
\affiliation{Korea Astronomy and Space Science Institute 776, Daedeokdae-ro, Yuseong-gu, Daejeon, Republic of Korea 305-348}

\author[0000-0001-5455-2546]{Heesu Yang}
\affiliation{Korea Astronomy and Space Science Institute 776, Daedeokdae-ro, Yuseong-gu, Daejeon, Republic of Korea 305-348}

\author[0000-0001-9982-2175]{Vasyl Yurchyshyn}
\affiliation{Big Bear Solar Observatory, New Jersey Institute of Technology, 40386 North Shore Lane, Big Bear City, CA 92314-9672, USA}

\author[0000-0002-7073-868X]{Jongchul Chae}
\affiliation{Astronomy Program, Department of Physics and Astronomy, Seoul National University, Seoul 08826, Republic of Korea}

\author[0000-0002-7737-740X]{Donguk Song}
\affiliation{National Astronomical Observatory of Japan, 2-21-1 Osawa, Mitaka, Tokyo 181-8588, Japan}

\author[0000-0001-9806-2485]{Maria S. Madjarska}
\affiliation{Astronomy Program, Department of Physics and Astronomy, Seoul National University, Seoul 08826, Republic of Korea}
\affiliation{Max Planck Institute for Solar System Research, Justus-von-Liebig-Weg 3, 37077, G\"ottingen, Germany}

\begin{abstract}
Light bridges (LBs) are relatively bright structures that divide sunspot umbrae into two or more parts. Chromospheric LBs are known to be associated with various activities including fan-shaped jet-like ejections and brightenings. Although magnetic reconnection is frequently suggested to be responsible for such activities, not many studies present firm evidence to support the scenario. We carry out magnetic field measurements and imaging spectroscopy of a LB where fan-shaped jet-like ejections occur with co-spatial brightenings at their footpoints. We study LB fine structure and magnetic field changes using TiO images, Near-InfraRed Imaging Spectropolarimeter, and H$\alpha$ data taken by the 1.6~m Goode Solar Telescope. We detect magnetic flux emergence in the LB that is of opposite polarity to that of the sunspot. The new magnetic flux cancels with the pre-existing flux at a rate of $5.6\times10^{18}$~Mx~hr$^{-1}$. Both the recurrent jet-like ejections and their base brightenings are initiated at the vicinity of the magnetic flux cancellation, and show apparent horizontal extension along the LB at a projected speed of up to {\kms{18.4}} to form a fan-shaped appearance. Based on these observations, we suggest that the fan-shaped ejections may have resulted from slipping reconnection between the new flux emerging in the LB and the ambient sunspot field.
\end{abstract}

\section{INTRODUCTION} \label{sec:intro}

Sunspot light bridges (LBs) are relatively bright and elongated structures that either protrude from one side of the penumbra into the umbra or completely divide a sunspot umbra into two or more parts. They are formed by the coalescence of umbrae in complex active regions or during the final stage of the active region evolution \citep{Bra64, Gar87}. Among various ways to classify LBs, their brightness and morphological properties are often used to separate LBs into two categories, strong or faint LBs. Strong LBs fully separate umbral cores and have brightness comparable to that of the penumbra \citep{Sob93, Sob94, Jur06, Rim08}. They very often display a granular structure with granules that are slightly smaller than the quiet-Sun granules \citep{Sob94}, and their formation is closely related to the sunspot decay \citep{Vaz73}. Faint LBs, composed of elongated bright grains, penetrate the umbra  \citep{Lit91, Sob93}. The size of these grains is comparable to those of umbral dots. Strong LBs can be further classified into two types: i) penumbral LBs that show a fine filamentary structure and ii) photospheric or granular LBs that show a fine structure similar to the photospheric granulation \citep{Rim97, Lag14}.

Compared to the magnetic field of the surrounding umbra, magnetic fields in LBs were found to be of the same polarity as the host umbra but of lower strength and more inclined \citep{Bec69,Lit90,Lit91,Rue95,Lek97,Jur06,Tor15}, supporting the idea of the field-free plasma intrusion into the sunspot umbra \citep{Spr06}. The azimuth of the LB magnetic field shows a tendency to be aligned with the LB orientation \citep{Jur06, Lek97}. \citet{Lek97} suggested a magnetic canopy structure above a LB based on the analysis of inverted vector magnetograms. Later \citet{Jur06} reported a stratification of the magnetic field strength and temperature above LBs, which is consistent with the idea of magnetic canopy.

The presence of inclined LB magnetic fields embedded in more vertical umbral field provides a favorable environment for magnetic reconnection to proceed at locations of strong discontinuity of the magnetic field. High resolution observations of strong LBs carried out by \citet{Ber03} revealed an enhanced brightness in TRACE 1600~\AA\ images, which indicates a magnetically heated LB chromosphere. The detection of persistent or transient brightness enhancements in the chromosphere above LBs suggests chromospheric heating in the LBs \citep{Lou08,Lou09,Tor15}.

In addition to brightness enhancement, several studies reported jet-like plasma ejections of LBs seen in various chromospheric \citep{Roy73, Asa01, Rob16,Shi09} and UV spectral lines \citep{Tor15, Tia18}. A typical property of the LB ejections is their intermittency. \citet{Asa01} reported the repetitive occurrence of H$\alpha$ surges with a mean lifetime of about 10~minutes. The surges showed a maximum apparent length of 20~Mm and a projection corrected mean velocity of \kms{40}. {LB jets often display fan-shaped appearance and their tips follow a parabolic trajectory indicating an impulsive upward acceleration affected by the gravitational force \citep{Tor15, Rob16, Bai19}.}

Based on the dynamical properties of LB ejections and the underlying magnetic fields it was suggested that magnetic reconnection between the LB field and the umbral field may be driving these ejections. However, it was pointed out by \citet{Tia18} that no solid evidence for LB jets driven by magnetic reconnection have been presented so far. {Only few studies detected mixed polarity fields in LBs, while most of the relevant studies failed to detect such fields.} \citet{Bha07} observed opposite polarity magnetic elements in a strong penumbral LB accompanied by Ellerman-Bomb like brightenings \citep[also see][]{Ell17, Rob16, Roy73} and dark plasma ejections. Because of the saturation of the magnetic field measurements used in this study, quantitative estimates of the associated flux cancellation could not be performed. {More recently, \citet{Yan19} detected an increase of the transverse magnetic field at the edge of an LB where a fan-shaped jet occurred spatially and temporally correlated with the emerging magnetic flux.} \citet{Bai19} analyzed Stokes V observations in a LB obtained by the Hinode/SP \citep{Ich08} and found abnormal three lobe profiles that generally suggest the presence of opposite magnetic polarities along the line-of-sight (LOS). \cite{Son17} reported photospheric signatures of an emerging magnetic field in an LB appearing prior to an inverted-Y shaped H$\alpha$ jet and argued that {these observations are a direct evidence} for magnetic reconnection occurring in an LB.

Taking advantage of the high spatio-temporal resolution of the Goode Solar Telescope (GST), we detected magnetic field emergence and cancellation in an LB. The emerging magnetic flux had a polarity opposite to that of the host sunspot field. Fan-shaped jet-like H$\alpha$  ejections and brightenings were detected recurrently appearing during a 4-hour observing period. These {observations} are unique as they allow us to perform both qualitative and quantitative analysis of the emergence and cancellation processes that were not fully addressed in previous studies.

\section{OBSERVATIONS AND DATA ANALYSIS}\label{sec:obs and data}

\begin{figure}[tb]
\begin{center}
    \includegraphics[width=0.98\textwidth]{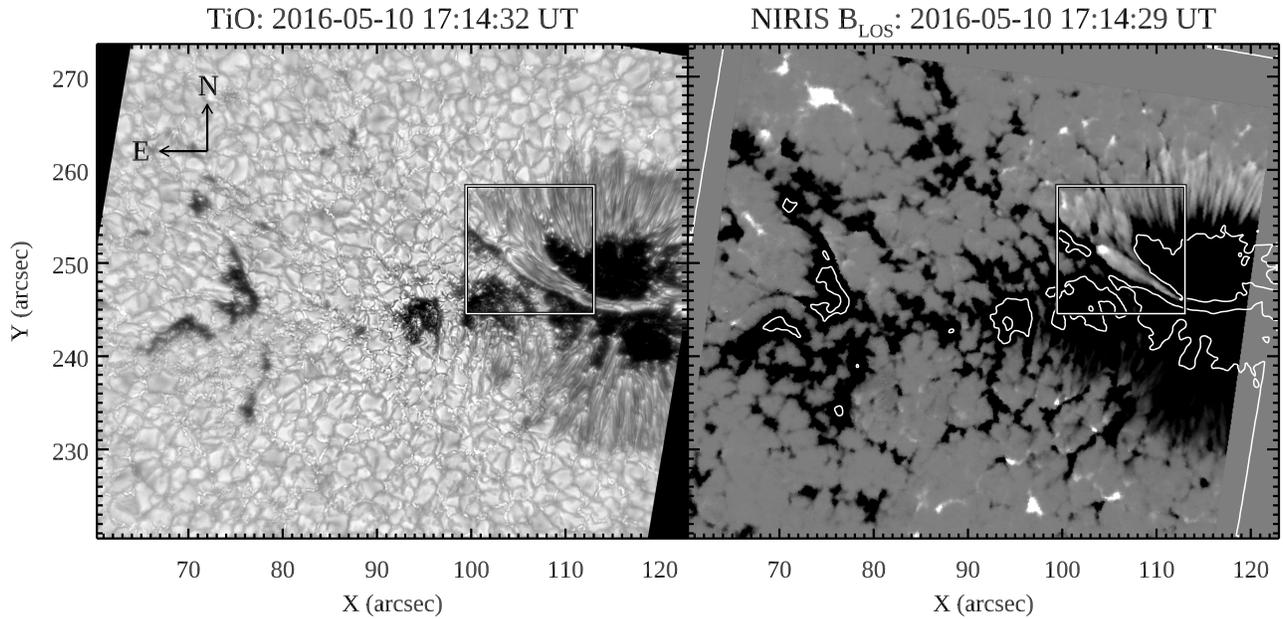}
     \caption{TiO intensity (left) and NIRIS longitudinal magnetic field map (right) of NOAA AR 12547. Solid square indicates the FOV of the cropped region, the same as the FOV of each panel in Figure~\ref{enlarge_view}. White contours in the right panel represents umbra-penumbra boundary defined as $0.69$ level of the mean TiO quiet region intensity.}\label{full_AR}
\end{center}
\end{figure}

{Our obervations started on 2016 May 10 at 16:08:00~UT.} At this time the NOAA active region 12542 was in its early phase of fragmentation and was located at heliocentric coordinates (110\arcsec, 245\arcsec) with heliocentric angle 16.2$\degr$ ($\mu = 0.96$). The sunspot was already in its mature stage with a fully developed penumbra and a number of strong light bridges (LBs) when it appeared on the eastern limb on May 4. As shown in Figure~\ref{full_AR}, the sunspot is of negative magnetic polarity, and its umbra is divided into four parts by three strong LBs. Two of the LBs lie along the north-south direction and have granular inner structure, while the third one is oriented along the east-west direction and has a penumbra-like appearance, specifically in the eastern part near the adjacent penumbra (106\arcsec, 248\arcsec). Compared to the other two granular LBs, this strong penumbral LB, which is the main target of this study, was accompanied by conspicuous chromospheric ejections and brightenings intermittently but continuously appearing throughout our observing time.

We observed the LB for about four hours starting from 16:08:22~UT on 2016 May 10 using GST with the aid of the adaptive optics system \citep{cao10a} under a good seeing condition. The broadband TiO Imager \citep{cao10b} acquired photospheric data every 15~s using a broadband (10~\AA) filter centered at 7057~\AA\, with a pixel scale of 0\arcsec.0375. The Visible Imaging Spectrometer \citep[VIS;][]{cao10a} produces a narrow 0.07~\AA\, bandpass over a 70\arcsec\ circular field of view (FOV) using a single Fabry-P\'{e}rot etalon. The chromospheric data were obtained at a number of selected wavelength positions along the H$\alpha$ spectral line. The obatined raster images have a pixel scale of 0\arcsec.029. Line scans were performed using various combinations of $\pm1.2$, $\pm1.0$, $\pm0.8$, $\pm0.4$, and 0.0~\AA\ positions depending on the type of the observed chromospheric activity. Consequently, the time cadence of the chromospheric data varies over the observing run from 12~s to 90~s (in case of the $\pm0.8$~\AA\, off-band data). All TiO and H$\alpha$ data were speckle reconstructed using the Kiepenheuer-Institute Speckle Interferometry Package \citep{wog08}.

The Near Infra-Red Imaging Spectropolarimeter \citep[NIRIS;][]{cao12} allows us to perform a full Stokes spectroscopic polarimetry using the Fe I 1565 nm doublet over a 85\arcsec\, round FOV with an aid of a dual Febry-P\'{e}rot etalon. Stokes I, Q, U, and V profiles were obtained every 72~s with a pixel scale of 0\arcsec.083. {A conventional Stokes inversion method developed by J. Chae \citep{Lan92} that fits observed profiles assuming an ME atmosphere was applied to the data after polarization calibration.} All GST data were co-aligned and derotated using SDO/HMI {\citep{Pes12}} data as reference. For the analysis of the LB magnetic field and the associated jet activity, we used TiO, H$\alpha\pm0.8$~\AA\, and NIRIS data simultaneously taken between 17:00~UT and 20:00~UT.

\section{RESULTS}
\subsection{PHOTOSPHERIC FINE STRUCTURES AND MAGNETIC FIELD IN THE LIGHT BRIDGE}\label{sec:result_mag}

\begin{figure}[tb]
\begin{center}
    \includegraphics[width=0.45\textwidth]{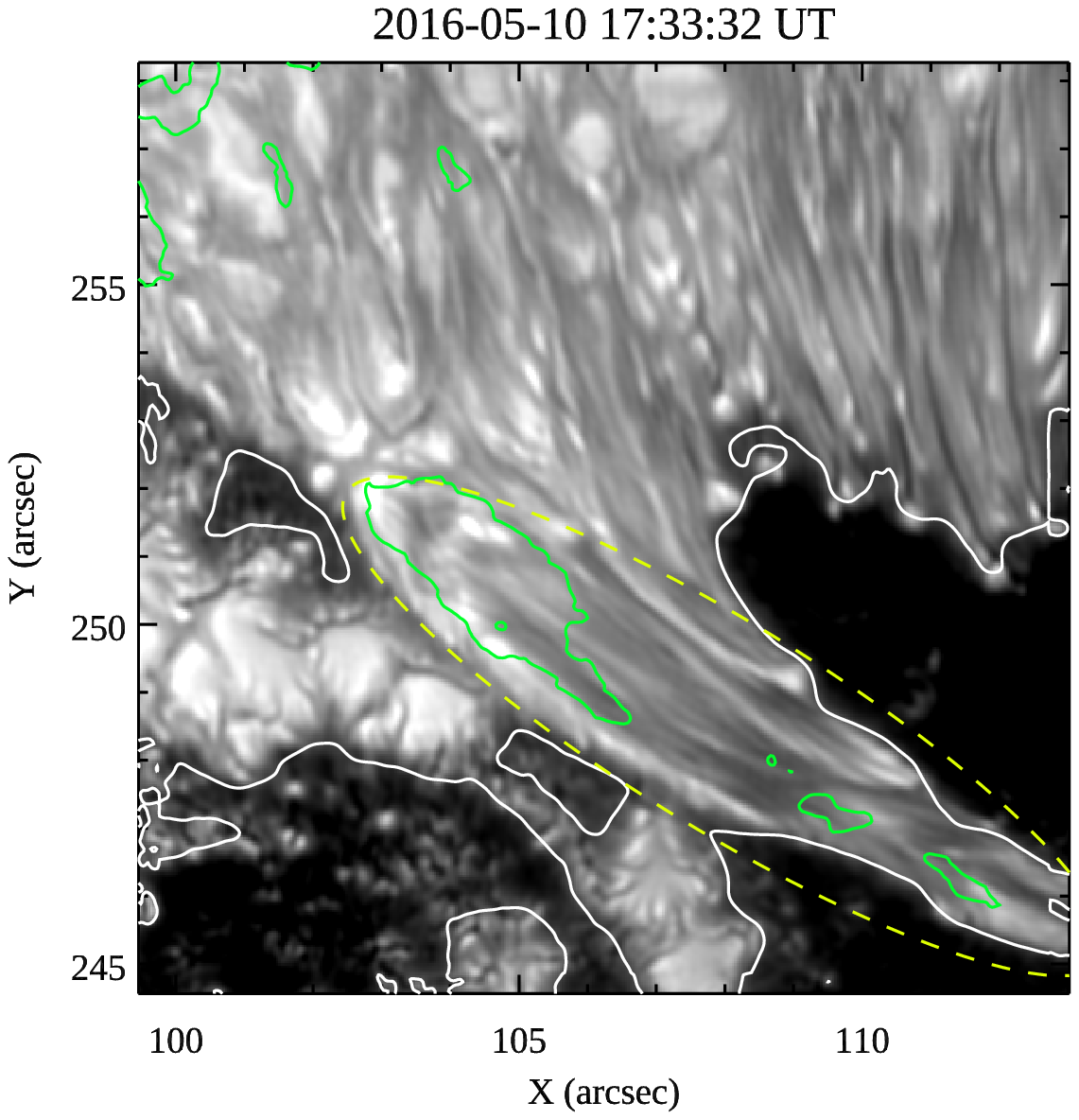}
    \includegraphics[width=0.45\textwidth]{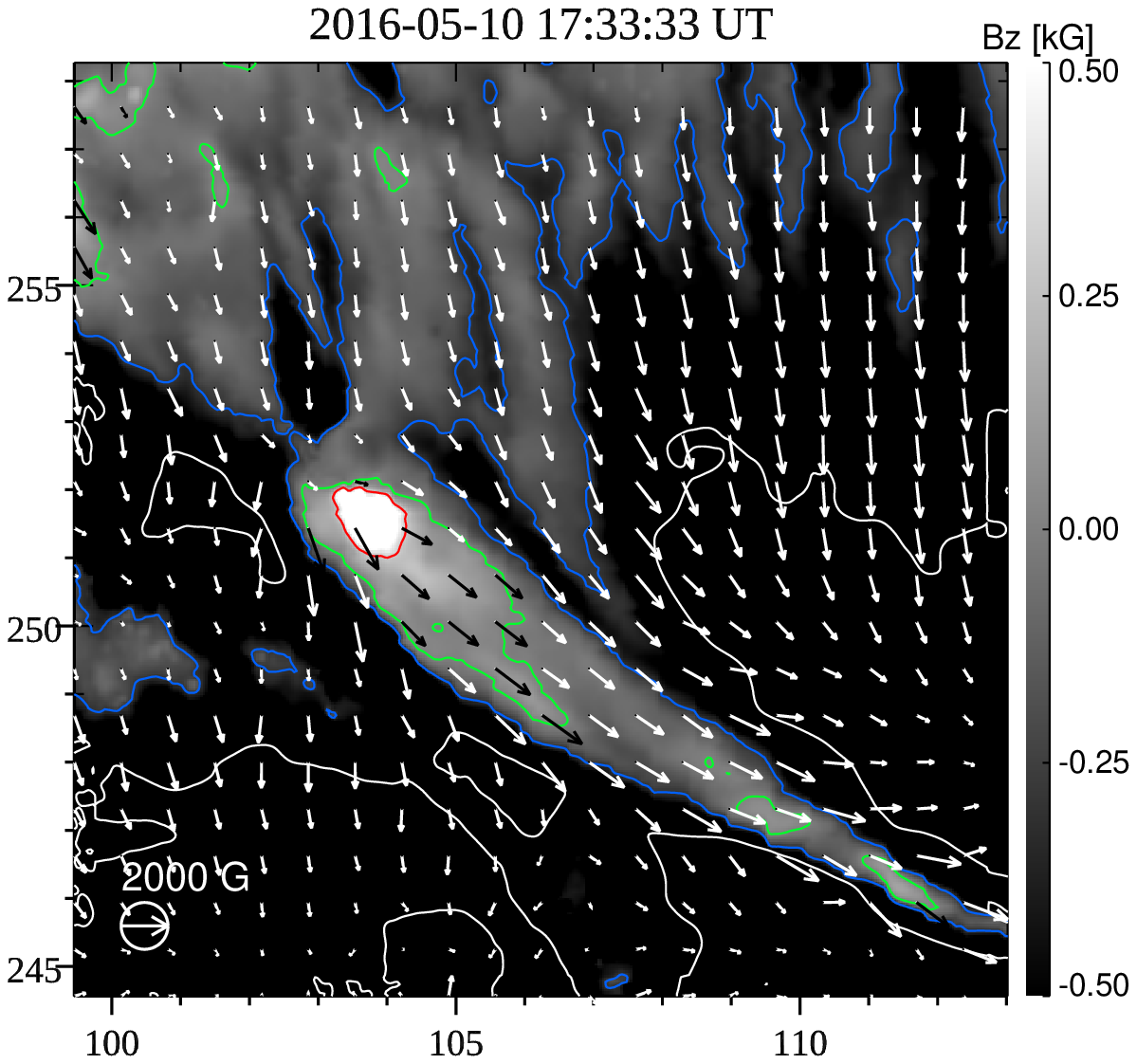}
    \includegraphics[width=0.45\textwidth]{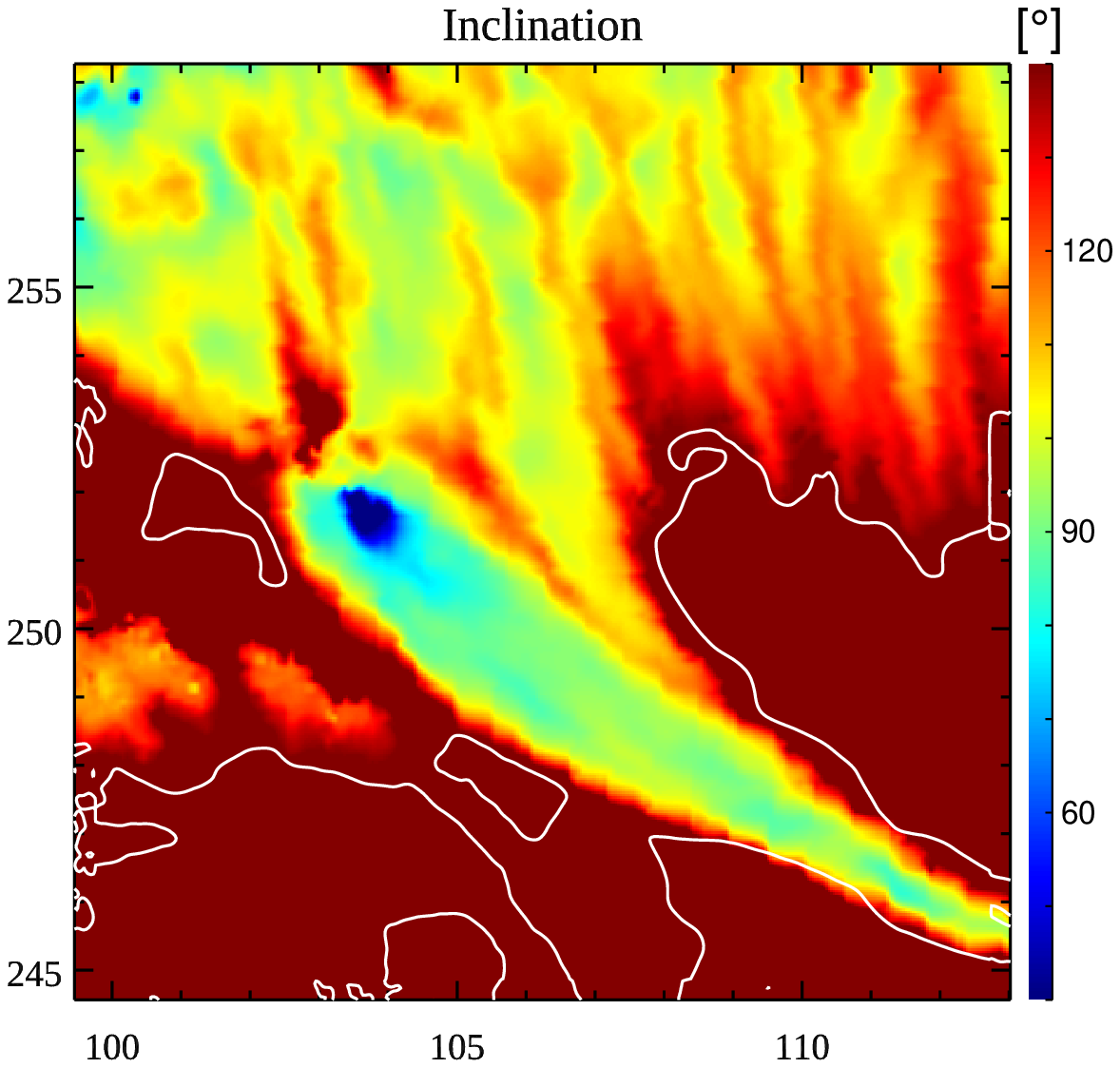}
    \includegraphics[width=0.45\textwidth]{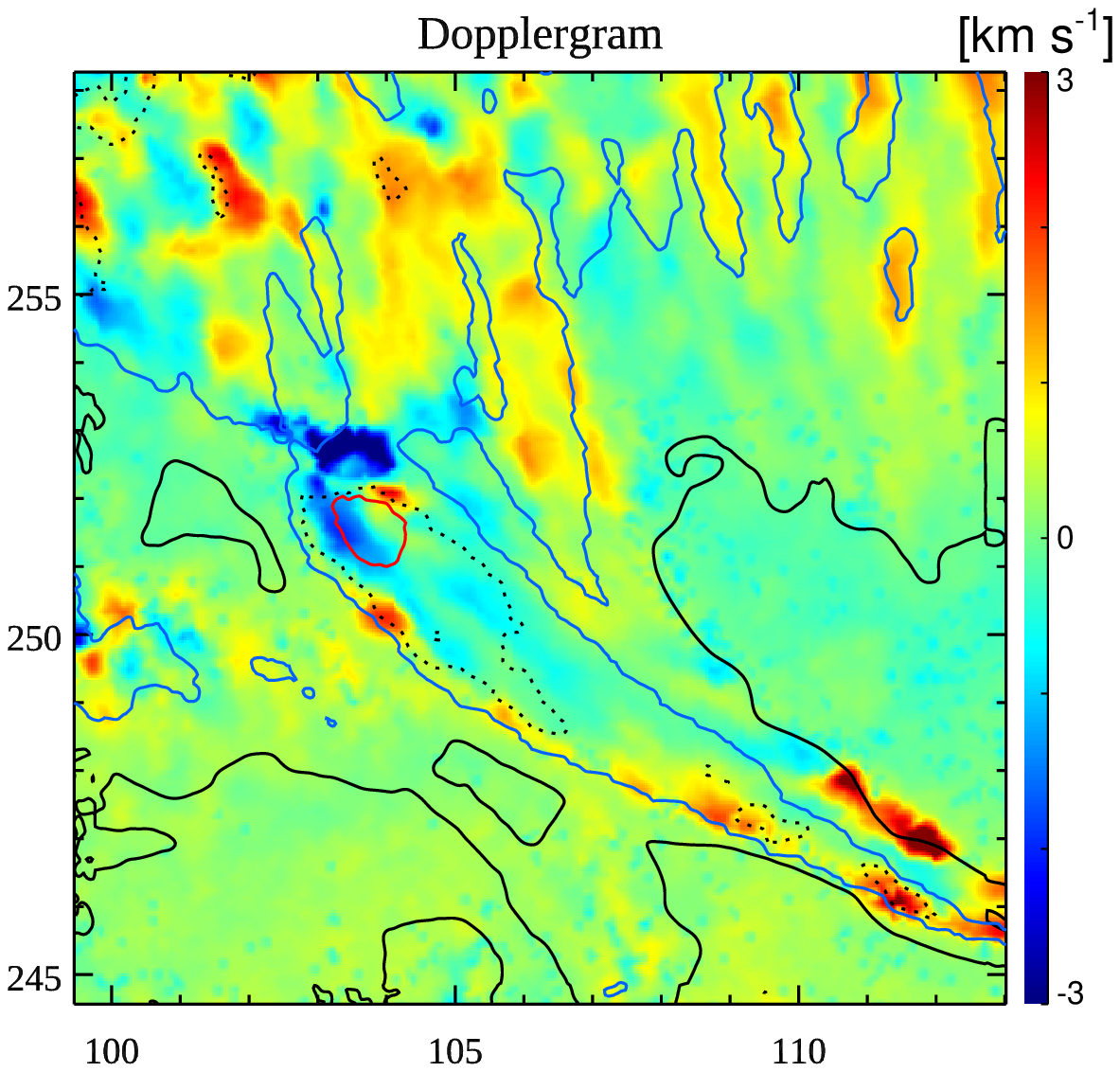}
     \caption{Enlarged view of interested LB region for different data sets. Top-left : TiO photospheric intensity map overlaid with $B_z$ contour at 0~G (green). Yellow ellipse outlines interested photospheric `{filament}-bundle'. Top-right : NIRIS vector magnetogram corrected for the projection effect. Vertical component of vector magnetic field $B_z$ (saturated at $\pm500$~G) in greyscale is overlaid with arrows representing transverse magnetic components $B_x$ and $B_y$. The length of an arrow corresponding to the diameter of the circle in the lower-left edge represents the field strength of 2~kG. $B_z$ contours are presented at $-300$~G (blue), 0~G (green), and 300~G (red). Bottom-left : inclination map saturated at $40\degr$ and $140\degr$. {Bottom-right : Doppler velocity map overlaid with $B_z$ contours at $-300$~G (blue), 0~G (dotted-black), and 300~G (red).} Umbra-penumbra boundary is outlined by white (or black for Dopplergram) contours. }\label{enlarge_view}
\end{center}
\end{figure}

Figure~\ref{enlarge_view} shows the area outlined by the white square in Figure~\ref{full_AR}. The fine structure of the LB is well pronounced in the TiO image (see the top-left panel). Contrary to the nearby granular LBs (lower-left quadrant of the top-left panel), the LB of interest in the center of the panel is composed of many dark and bright {threads},{ which we call LB filaments by analogy with penumbral filaments. Most of these LB filaments look} very similar to the uncombed filamentary structure of the sunspot penumbra. This similarity makes the LB look like a continuation of the adjacent penumbra with the {filaments} oriented predominantly along the axis of the LB. Most of the {filaments} appear to begin near the location with x and y heliocentric coordinates of 104\arcsec, 251\arcsec, respectively. Their tails are within the umbra at either edges of the LB curling toward the umbra. As a result, the assembly of bright and dark {filaments} forms a single elongated ``{filament} bundle'' of a ~8\arcsec\ lengt and ~3\arcsec\ width, with a well-defined head at (104\arcsec, 251\arcsec), and a relatively scattering tail (outlined by the yellow ellipse in the top-left panel of Figure~\ref{enlarge_view}).

The top-right panel of Figure~\ref{enlarge_view} shows the vector magnetogram obtained by applying the ME inversion to the NIRIS Stokes profiles. From the comparison of the TiO images and the vector magnetogram, we can confirm that the LB {filaments} follow the direction of the transverse magnetic field in the LB \citep{Lek97}. One important finding is that the magnetic field in the LB is positive, while the host sunspot magnetic field is negative. Since the vector magnetogram was corrected for the projection effect, the $B_z$ outlined by the green contour has a real positive component. The vertical field strength is relatively weak, ranging from 0~G to 300~G, while the transverse field strength is around 1200~G nearly everywhere in the LB, except for the location of the vertical magnetic field concentration at (104\arcsec, 251\arcsec). This corresponds to the location of the head of the {filament} bundle in the TiO image (top-left). The inclination map obtained from the projection-corrected vector magnetogram (bottom-left) shows that the magnetic field in the LB region is predominantly horizontal ($80\degr$--$100\degr$). At the same time, the relatively strong positive $B_z$ component is concentrated at the {filament} bundle at the umbra-penumbra boundary. The combination of the vector magnetogram and the inclination map suggests that the magnetic field along the LB may represent a low-lying arcade having one-side of the footpoints in the positive flux concentration and the other in the nearby umbra. As mentioned above, the LB magnetic field may not have a typical bipolar configuration as the negative magnetic polarity footpoint appears to be scattered into the umbra. The comparison between the TiO and NIRIS data reveals that the {LB filaments} are proxies of the transverse magnetic field thus providing detailed information on the field connectivity, including the footpoint locations.

Doppler velocities $v_D$ obtained from the ME inversion of the NIRIS data and with the mean velocity of the umbrae subtracted (bottom-right panel in Figure~\ref{enlarge_view}), show downflows with speeds of about \kms{3} (red) located in the vicinity of the positive flux concentration and the outer boundary of the LB. Except for the regions with significant downflows, the rest of the LB shows slightly upflow (blueshifted pattern) signatures ($v_D$ of up to \kms{-0.8}) relative to the umbral mean velocity. Although the upflow speed of \kms{0.8} is relatively low, such a $v_D$ pattern with a blueshift in the middle and a redshift at the footpoints is consistent with the observations of $\Omega$ loops emerging through the photosphere.

\begin{figure}[tb]
\begin{center}
    \includegraphics[width=0.96\textwidth]{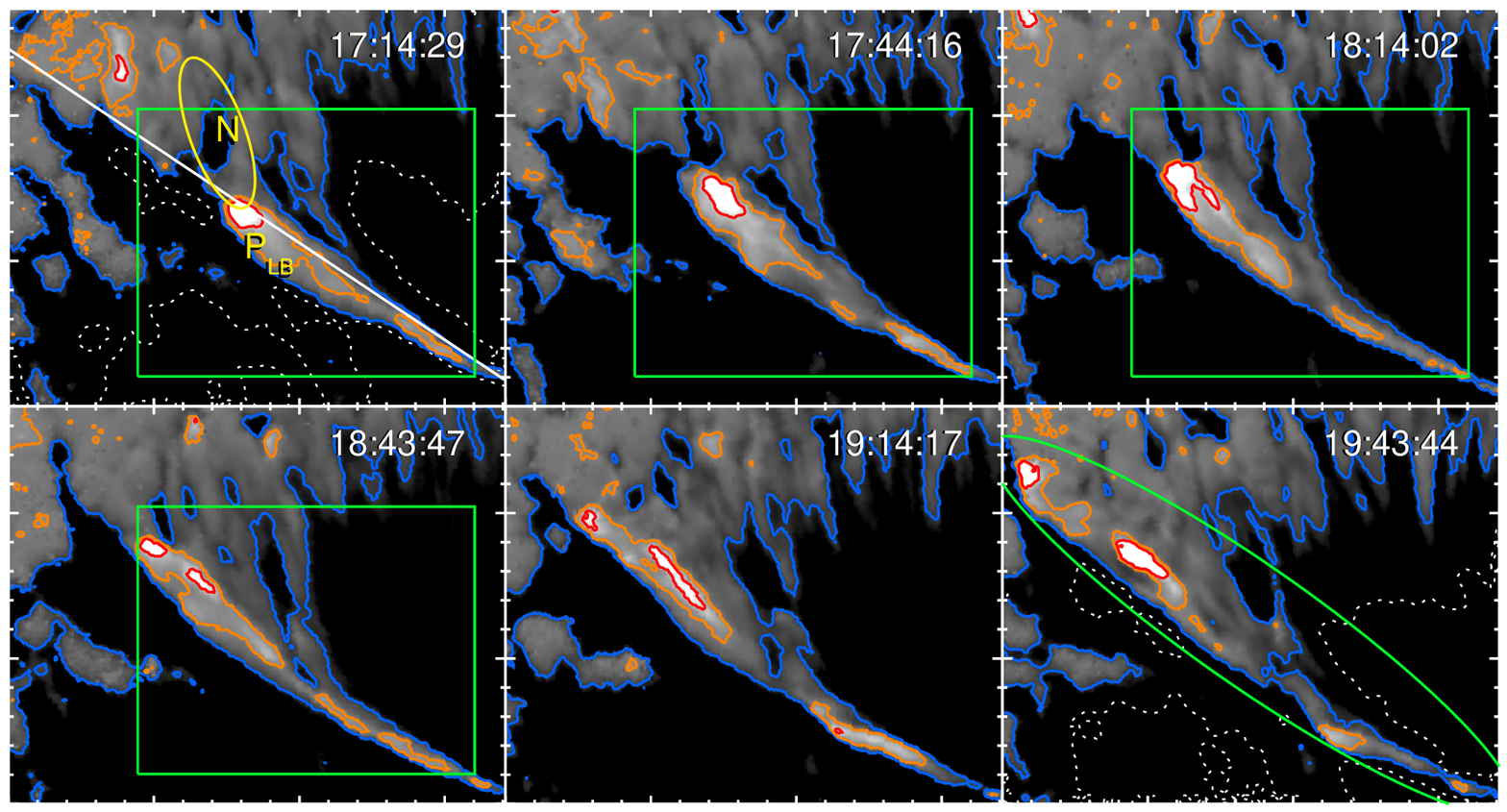}
    \includegraphics[width=0.96\textwidth]{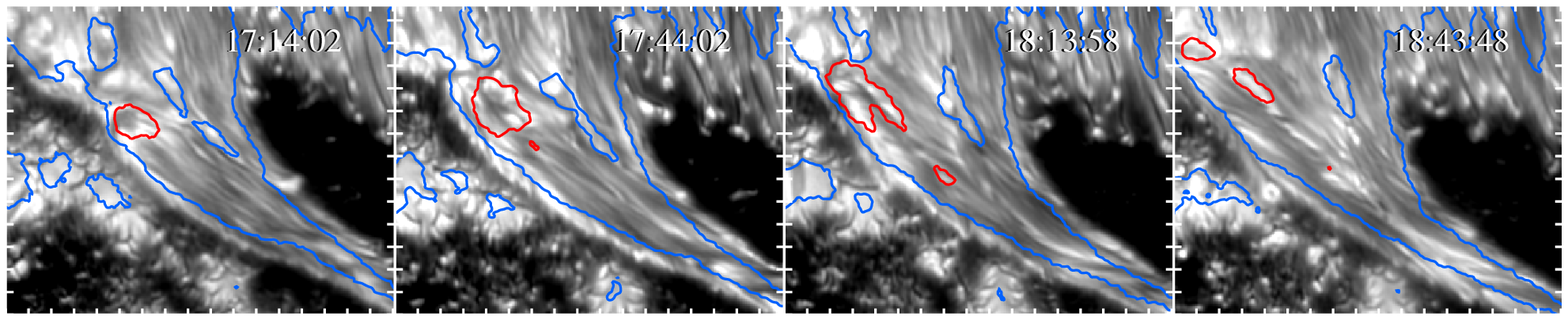}
    \caption{Top: time series of $B_z$ at every 30 minutes during 2.5 hours saturated at $\pm500$~G overlaid by contours of field strength at $-300$~G (blue), $+5$~G (orange) and $+300$~G (red). {White dotted contours indicate umbra-penumbra boundary.} White line indicates the slit position along which the space-time (x-t) diagram was obtained. The positive magnetic {field} in the LB is labels as P$_{\text{LB}}$ and the pre-existing adjacent negative {field} as N in yellow color. {Magnetic flux of the LB is defined from the area marked by the bigger green ellipse, and that of N, for the area marked by the smaller yellow ellipse.} Bottom: simultaneously taken TiO intensity maps within a cropped FOV represented by a green rectangle overlaid by the same contours as in the top panels. }\label{bz_series}
\end{center}
\end{figure}

\begin{figure}[tb]
\begin{center}
    \includegraphics[width=0.96\textwidth]{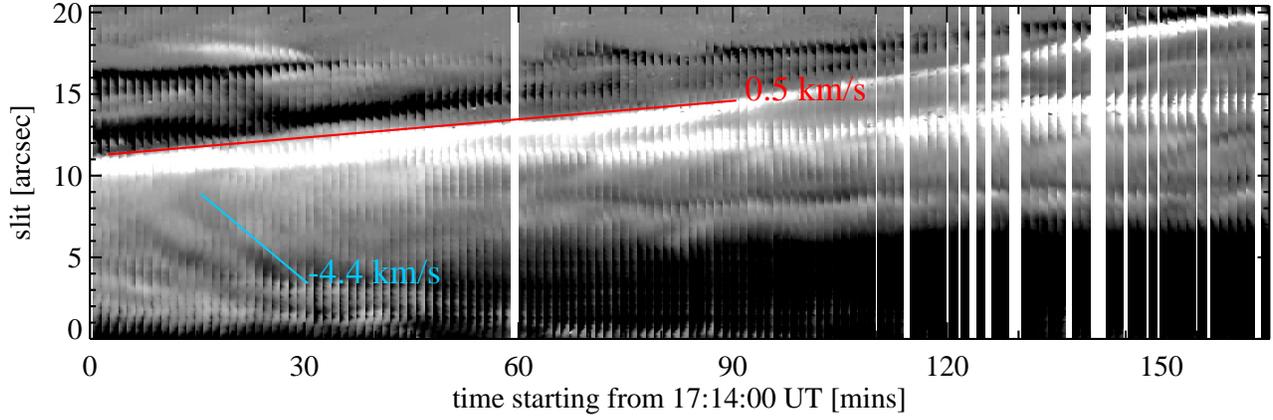}
    \caption{X-t diagram of the $B_z$ for 2.5 hours obtained along the slit shown in the top panel of Figure~\ref{bz_series}. Red and cyan curves represent the linearly fitted tracks of `P$_{{\text{LB}}}$' and the `opposite flow' of magnetic features mentioned in the text, respectively. The resulting velocity of each track is also presented in the same color as that of linear-fit curve, \kms{0.5} for the red curve and \kms{-4.4} for the cyan, respectively. }\label{xt_bz}
\end{center}
\end{figure}

Indeed, we found evidence for the emergence of positive magnetic flux in the LB. We also obtained a quantitative temporal magnetic flux profile, showing the emergence and the cancellation of the positive flux (Figure~\ref{bz_series}). A series of $B_z$ magnetograms shows that the positive magnetic flux P$_{\text{LB}}$ slowly and continuously moves in the north-east direction, while the apparent size of the flux patch as outlined by 300~G contour (shown with red curve in Figure~\ref{bz_series}) gradually increases until it encounters the adjacent negative flux, N, at around 18:00~UT. After that, P$_{\text{LB}}$ starts to decrease and continues to move forward. The apparent speed of P$_{\text{LB}}$ was estimated from the space-time (x-t) diagram to be \kms{0.5} (Figure~\ref{xt_bz}, red curve). The x-t diagram was obtained along the white slit shown in the first panel of Figure~\ref{bz_series}. Figure~\ref{xt_bz} also reveals the presence of a flow of magnetic features in the opposite direction to that of P$_{\text{LB}}$ during the early stage of the magnetic flux emergence (cyan curve). The apparent speed of this opposite flow is estimated from the x-t diagram to be \kms{4.4}. This flow was detected only in the early stage of emergence may be {because it was not parallel to the slit direction.} We set the slit orientation to best track the positive flux, {P$_{\text{LB}}$}. In combination with the upflow signatures seen in the $v_D$ map (Figure~\ref{enlarge_view}), this diverging flow may indicate magnetic flux emergence in the LB.

The bottom panel of Figure~\ref{bz_series} demonstrates the lengthening of the {filament} bundle along the LB as a result of the diverging flow. It is now well accepted that emerging magnetic flux is accompanied by a photospheric counterpart, such as elongation of granular-like structures along the transverse magnetic field direction \citep{Lim11}. \citet{Son17} also reported a similar pattern in a LB expanding at a rate of $\sim$\kms{2}, and interpreted it as a manifestation of an emerging flux. Figure~\ref{bz_series} also shows that the appearance of the observed LB filaments is quite similar to that of typical penumbral filaments. Interestingly, the observational properties of the LB filaments studied here that support the idea of an emerging low-lying arcade, are also consistent with some of the observational properties of penumbral filaments.{ Those properties include the presence of bright penumbra grains, horizontal magnetic fields, and downflows at the filament tails \citep[][and references therein]{Tiw13}.}

\begin{figure}[tb]
\begin{center}
    \plotone{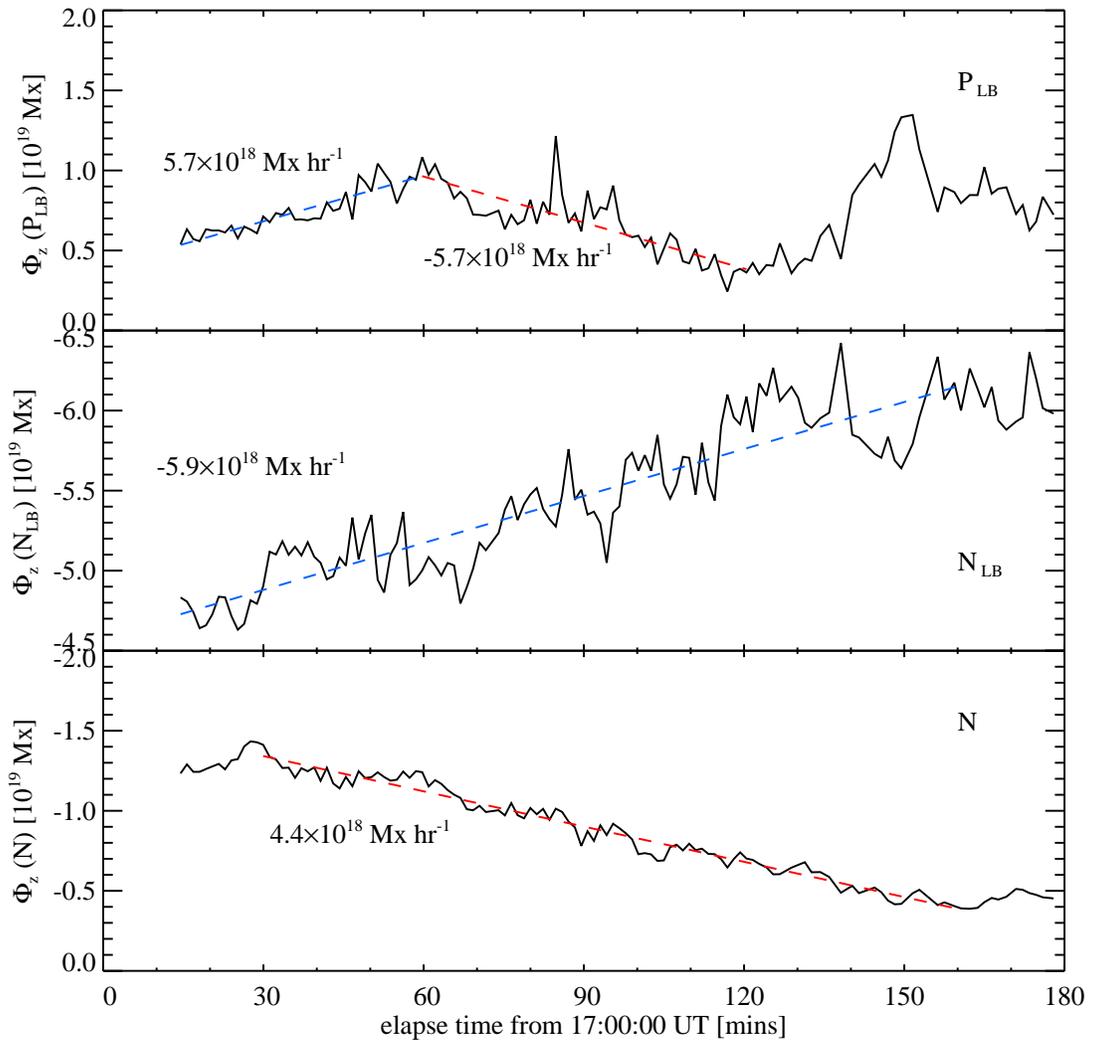}
    \caption{Temporal change of the positive {and the negative} magnetic flux in the LB, $\Phi_z(\text{P}_{\text{LB}})$ (top), {$\Phi_z(\text{N}_{\text{LB}})$ (middle)}, and the adjacent, pre-existing negative magnetic flux, $\Phi_z$(N) (bottom). }\label{flux_plot}
\end{center}
\end{figure}

{We measured the temporal change of both the emerging magnetic flux in the LB and the adjacent negative flux. To extract the LB magnetic field, both the positive and the negative polarities, we defined an elliptic region (green ellipse shown in Figure~\ref{bz_series}) as large as the size of the elongated filament bundle but small enough to exclude any strong positive flux from outside of the LB region. In order to exclude strong umbral magnetic field from the elliptic region, we masked out the magnetic field within the region where the TiO intensity is lower than 0.69 level of the mean quiet region. We then integrated the vertical magnetic component, $B_{\text{z}}$, stronger than 5~G over pixels where the inclination angle is smaller than $90\degr$ (Figure~\ref{flux_plot}, top panel) to compute the positive magnetic flux P$_{\text{LB}}$. The negative magnetic flux in the LB, N$_{\text{LB}}$ was obtained in the same way as P$_{\text{LB}}$ by integrating the $B_{\text{z}}$ component stronger than $-5$~G and weaker than $-1000$~G within the same elliptic region (Figure~\ref{flux_plot}, middle panel). We note that it is difficult to distinguish the negative magnetic field of the LB from the umbra when they have the same magnetic polarity. We assume that a magnetic field weaker than $-1000$~G (comparable to the maximum strength of P$_{\text{LB}}$) represents the negative magnetic field of the LB. The magnetic flux of the adjacent negative field N was measured within a small elliptic region (yellow ellipse shown in Figure~\ref{bz_series}) and displayed in the bottom panel of Figure~\ref{flux_plot}. The negative $B_{\text{z}}$ component stronger than $-5$~G was integrated within the elliptic region.}

{Figure~\ref{flux_plot} shows both the emergence and cancellation of the LB magnetic field. We find that the positive magnetic flux in the LB increases for the first $\sim45$ minutes, during which we observed enlargement of both the photospheric filament bundle and the positive flux patch (Figure~\ref{bz_series}). We can see that the LB negative flux also increases during this time period at a comparable emergence rate ($5.7\times10^{18}$\,Mx~hr$^{-1}$ for P$_{\text{LB}}$ and $5.9\times10^{18}$\,Mx~hr$^{-1}$ for N$_{\text{LB}}$). After around 18~UT, when the P$_{\text{LB}}$ encountered the adjacent negative field, N, the positive flux begins to decrease, which lasted until about 19~UT. The bottom panel of Figure~\ref{flux_plot} shows that the magnetic flux of N also decreases during this time period. We confirm that the effect of inflow or outflow of the positive magnetic flux across the boundary of the selected region is negligible. Note that both the negative magnetic flux of N$_{\text{LB}}$ and N show gradual increase or decrease during the observational time, while the positive magnetic flux of P$_{\text{LB}}$ shows increase-followed-by-decrease patterns. This indicates that the LB magnetic field kept emerging, and at the same time the positive magnetic field experienced magnetic cancellation with the pre-existing sunspot magnetic field N. Assuming that the decrease of the observed flux was purely due to magnetic cancellation, the cancellation rate during the time interval between 18:00~UT and 19:00~UT is estimated to be {$5.7\times10^{18}$\,Mx~hr$^{-1}$} for P$_{\text{LB}}$ and $4.4\times10^{18}$\,Mx~hr$^{-1}$ for N. The constantly evolving and elongating photospheric filament bundle also supports the idea of continuous magnetic flux emergence.}

\subsection{FAN-SHAPED JET-LIKE EJECTIONS}\label{sec:result_jet}
\begin{figure}[tb]
\begin{center}
    \includegraphics[height=0.7\textheight]{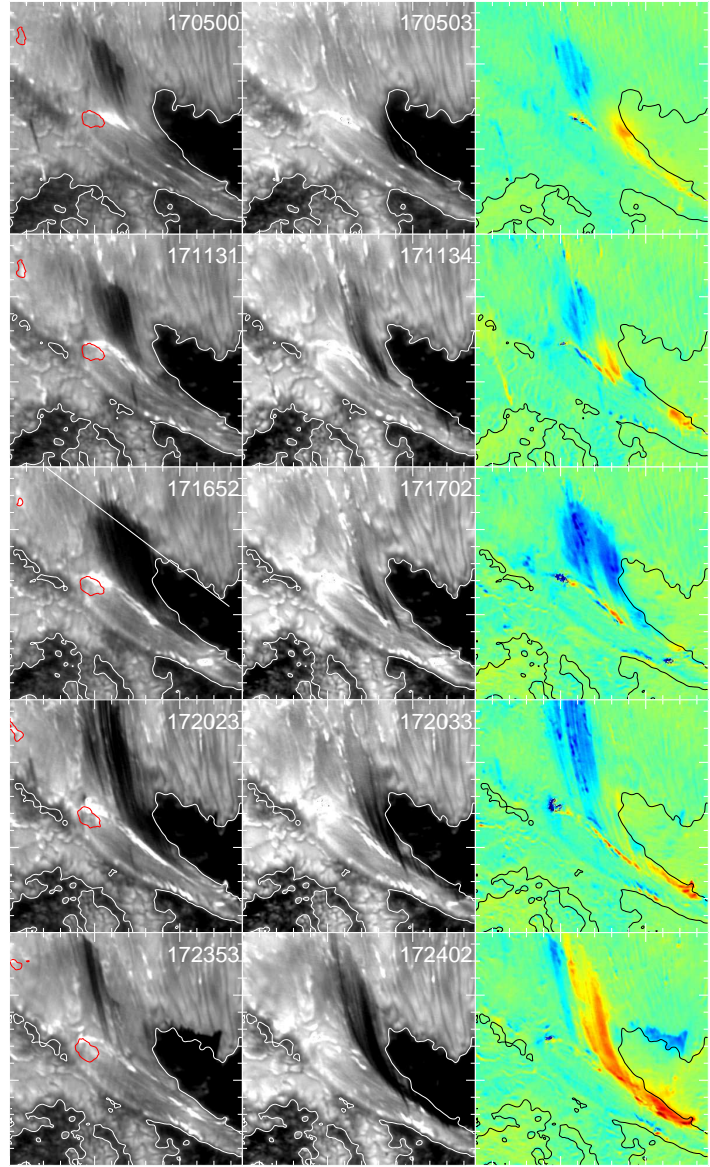}
    \caption{Time series of H$\alpha-0.8$\AA\, (left), H$\alpha+0.8$\AA\, (middle), and Dopplergram (right) during one episode of jet activity. Blue-shift (red-shift) components are colored blue (red) in the Dopplergram. White (resp. black) contours on the H$\alpha$ (resp. Dopplergram) represent umbral boundary obtained from the coaligned TiO data. Red contours on the H$\alpha-0.8$\AA\, data represent the positive magnetic flux at the field strength of 300~G. {White line in the 17:16:52~UT panel in the left column indicates the slit along which the x-t diagram (Figure~\ref{vis_xt17}) was obtained.} Each X- and Y-tick marks every 1\arcsec\, interval. Starting time for this episode is 17:03:04~UT and ending time is 17:26:50~UT. An animated version of this figure is available. It covers 26 minutes starting at 17:00:28~UT, and the realtime duration is 3 seconds. Times presented in each panel are in UT.}\label{vis_time_series}
\end{center}
\end{figure}

The chromospheric observations revealed intermittent but continuous jet-like ejections with brightenings at the base occurring along the LB. A bundle of jet strands aligned with the long axis of the LB was reliably observed in the H$\alpha$$-0.8$~\AA\, data (left column of Figure~\ref{vis_time_series}). Such ejections are often called fan-shaped jets (FSJs) due to their apparent shape. At the same time, a prominent elongated brightening was also visible at the base of the jet strands. These FSJs appeared intermittently but continuously throughout our observing time, mostly at the same location. Once a FSJ started, it lasted for about 20~minutes only to be promptly replaced by another FSJ event.

\begin{figure}[tb]
\begin{center}
    \includegraphics[height=0.5\textheight]{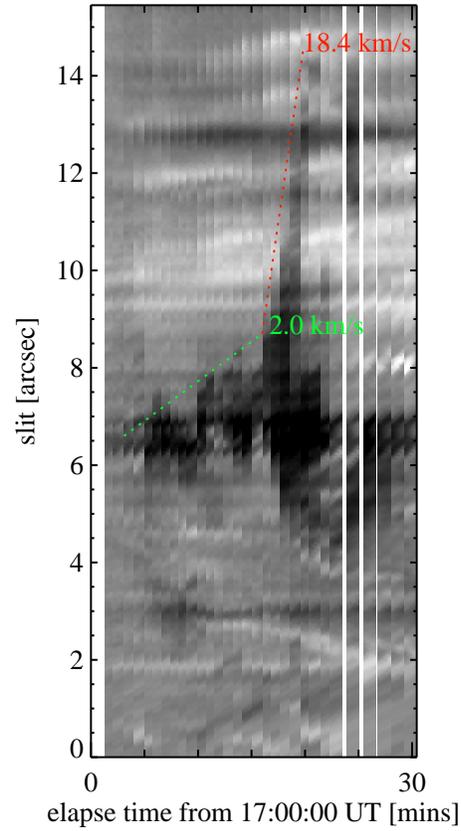}
    \caption{Space-time plot of H$\alpha-0.8$\AA\, obtained along the white line in Fig~\ref{vis_time_series}. Overplotted dotted lines indicate linear fits of the front of a dark structure that indicate a trace of the fan-shaped jet (FSJ) strand.}\label{vis_xt17}
\end{center}
\end{figure}

From H$\alpha$ data, we found three interesting properties of the FSJs. First, they originate at the location of the positive flux emergence alongside with an intense H$\alpha$ brightpoint in their base. Second, the FSJs show apparent horizontal broadening with time. Third, the FSJs' broadening is accompanied by an elongated H$\alpha$ brightening at their base. {This elongated brightening is not co-spatial with the initial brightpoint.} The left column of Figure~\ref{vis_time_series} shows the time series of the H$\alpha$$-0.8$~\AA\, data for about 20 minutes covering one of the FSJ events. A dark plasma ejection with a relatively small width ($\sim2$\arcsec) was detected at 17:05~UT with an intense base brightening very close to the positive flux patch (red contour). A few minutes later, the ejection became wider as the FSJ strands and the base brightening extended along the LB toward the umbra (at 17:11:31~UT). This horizontal extension continued until the FSJ width reached its maximum at 17:16:52~UT, after which the width decreased, and only a thin strand at the initial position remained visible at 17:23:53~UT. The same behavior is also visible in the co-temporal H$\alpha$$+0.8$~\AA\, data (middle column), although the FSJ development appears to be temporally delayed. The FSJ reached its maximum width in the H$\alpha$$+0.8$~\AA\, close to 17:24:02~UT, approximately 8 minutes later than the maximum width time in the H$\alpha$$-0.8$~\AA\,. The apparent extension speed of the rightmost edge estimated from the x-t diagram (Figure~\ref{vis_xt17}) is {\kms{2.0}} at firt. Then the speed suddenly increased to {\kms{18.4}} at around {17:16:00~UT}. The diagram was generated using the white slit shown in the panel at 17:16:52~UT (Figure~\ref{vis_time_series}, left column).

{The pseudo-Dopplergram in the right column of Figure~\ref{vis_time_series} produced by subtracting H$\alpha$$+0.8$~\AA\, from H$\alpha$$-0.8$~\AA\,, shows that the FSJ appears slightly shifted south-west in the H$\alpha$$+0.8$~\AA\, compared to the FSJ seen in H$\alpha$$-0.8$~\AA\,.} Some of the red-shifted jet strands seem to anchor in the lower umbra, while the blue-shifted jet strands are rooted in the upper {umbra}.
Such spatial inconsistency between the off-band components was also reported in previous study by \citet{Rob16}. Some of the strong ejections detected in the red-wing from the lower boundary of the LB are investigated in detail by \citet{YanH19}.

\begin{figure}[tb]
\begin{center}
    \includegraphics[width=0.85\textwidth]{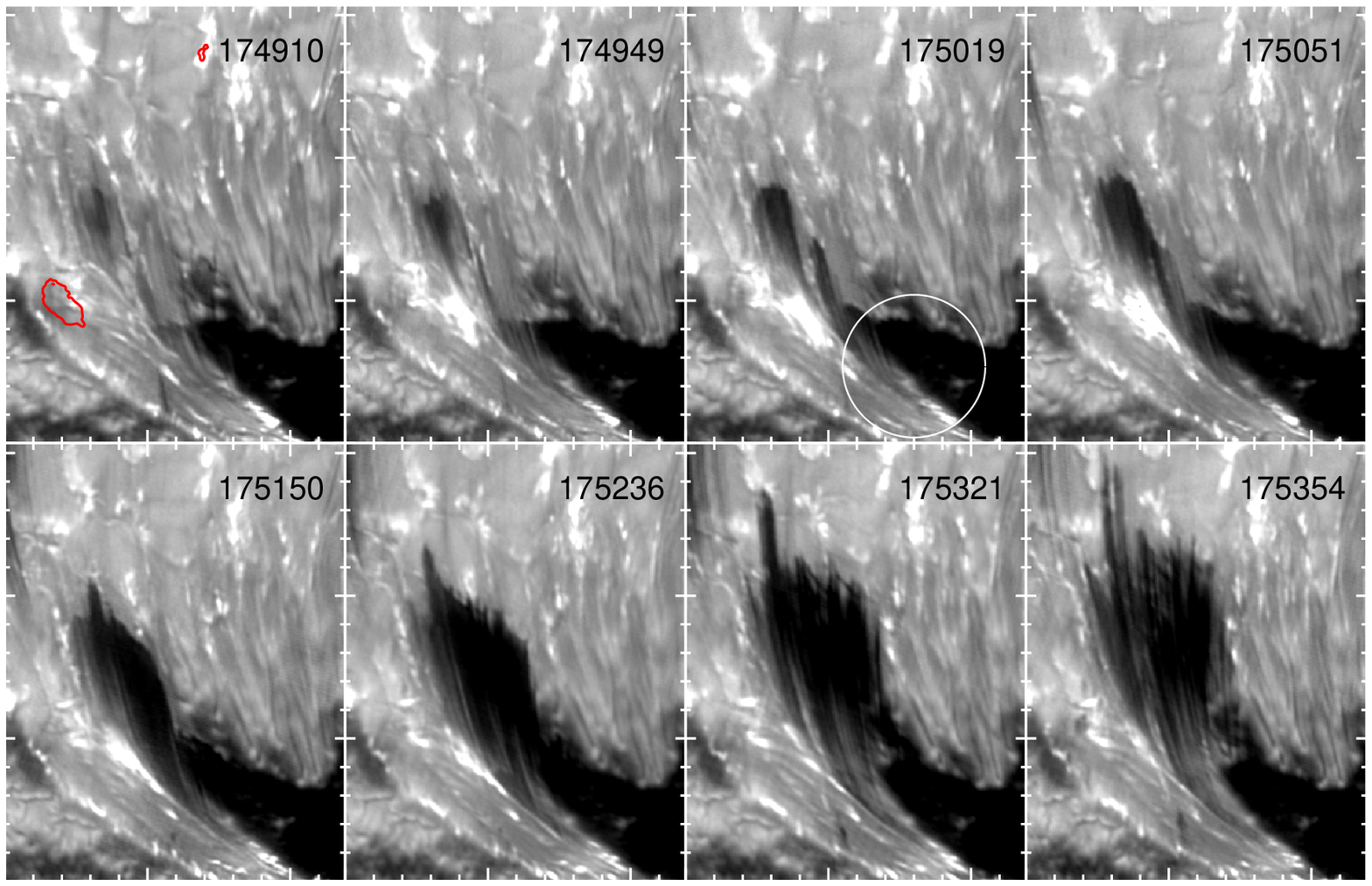}
    \includegraphics[width=0.85\textwidth]{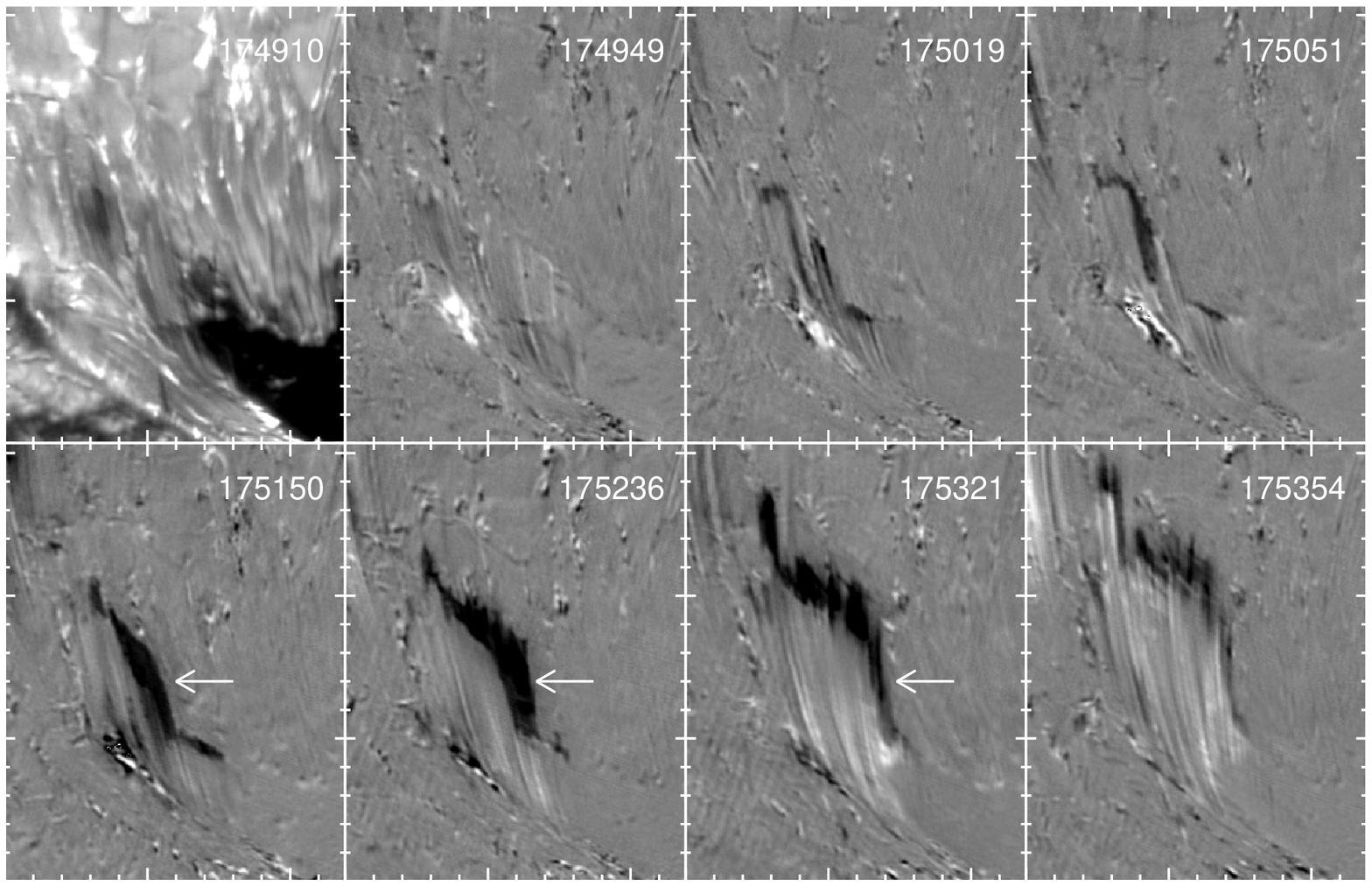}
    \caption{Early phase of fan-shaped jet cycle starting at 17:49:10~UT taken at H$\alpha-0.8$\AA\, (top) and the corresponding running difference images taken by subtracting data from the previous time from the data taken at displayed time (bottom). Each X- and Y-tick marks every 1\arcsec\, interval.}\label{vis_rd}
\end{center}
\end{figure}

Figure~\ref{vis_rd} shows another episode of the FSJs from the same location. A series of H$\alpha$$-0.8$\AA\, images at the early phase of the ejection shows two properties. One is that the lower part of the jet strands is highly curved toward the LB: the strands are nearly horizontal to the surface near the footpoint, while they become mostly vertical near the location of the elongated brightening (see the area inside the circle in the 17:50:19~UT panel). The significance of this is that the H$\alpha$ brightening may not be located at the footpoint of FSJs, but rather at the location where field lines turn from horizontal to vertical, in other words, at the beginning of the magnetic canopy. The other interesting property of the FSJ is that the jet strands horizontally drift towards the center of the umbra, while being straightened at the same time. The lower panels in Figure~\ref{vis_rd} show running difference images generated from the images in the top row.
The images were carefully co-aligned and normalized by making sure signatures of the umbra and penumbra are not present in the resulting running difference images. Newly appearing features are dark in the running difference images. The data show that the jet strands extend in length, indicating an upward ejection as well in horizontal direction toward the umbra (white arrow). The comparison of the fine {filaments} in the images at 17:53:54 and 17:50:19~UT indicates that the jet strands closer to the umbra are more vertical than those closer to the LB.

\begin{figure}[tb]
\begin{center}
    \includegraphics[width=0.95
    \textwidth]{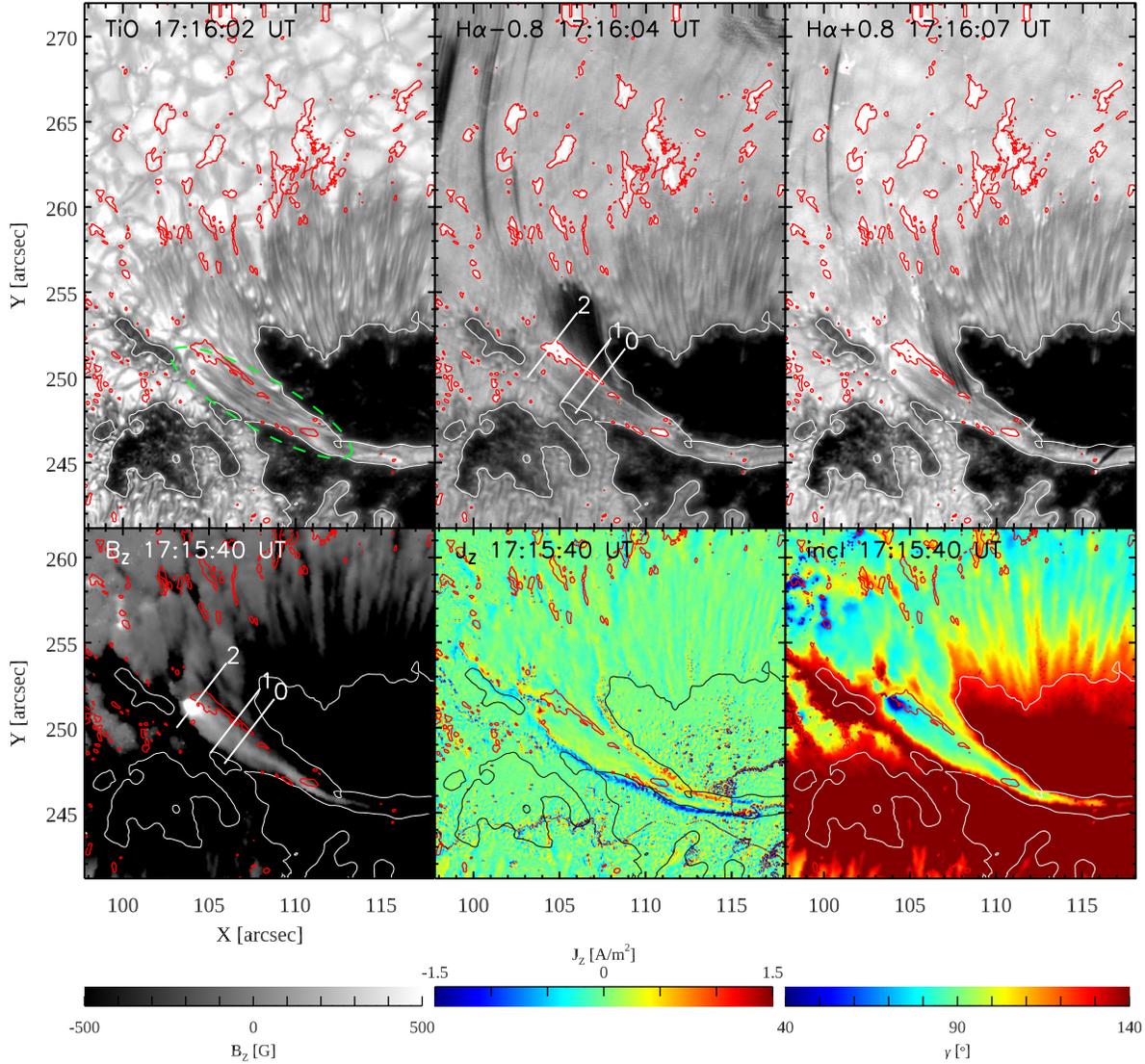}
     \caption{TiO, H$\alpha$ -0.8\AA\,, H$\alpha +0.8$\AA\,, B$_z$, J$_z$, and the inclination map taken at the time when the interested FSJ with the base brightening was firstly detected during our observing period. White contours represent $69\%$ level of the quiet region intensity, that roughly outline the penumbra-umbra boundary and umbra-LB boundary. Red contours represent the brightenings ($120\%$ of quiet region intensity) detected from H$\alpha -0.8$\AA\, data. White lines with numbers 0, 1, and 2 shown in the top-middle and bottom-left panels represent artificial slits along which profiles in Figure~\ref{profiles} are obtained. Vertical magnetic field, B$_z$ and the vertical current density, J$_z$ are saturated at $\pm500$ G and 1.5~$\textrm{A~m}^{-2}$, respectively. An animation of this figure is available. It covers one hour from 17:00:31~UT to 17:59:43~UT, and the realtime duration of the video is 7 seconds. All times are in UT.}\label{tio_vis_niris}
\end{center}
\end{figure}

In Figure~\ref{tio_vis_niris} we show the observed FSJ at the time of its maximum width along with a co-temporal TiO photospheric image and a NIRIS magnetogram. The umbra-penumbra and LB-umbra boundaries as well as H$\alpha$$-0.8$~\AA\ brightening are outlined by white (black in case of the $J_z$ map) and red contours, respectively. The brightening at the base of the FSJ is at the outer edge of the elongated photospheric {filament} bundle (TiO panel), which is the boundary between the emerging positive flux and the western umbral core ($B_z$ panel). We computed the vertical component of the electric currents ($J_z$, lower middle panel saturated at $\pm1.5~\textrm{A~m}^{-2}$) using the vector magnetic field data. Although both $B_z$ (lower left panel) and the inclination map (lower right) show that the polarity inversion line is co-spatial with the H$\alpha$ brightening, there is no strong vertical currents associated with that location. Instead, strong current elements are found at the boundary between the LB and the umbral cores.

\begin{figure}[tb]
\begin{center}
    \plotone{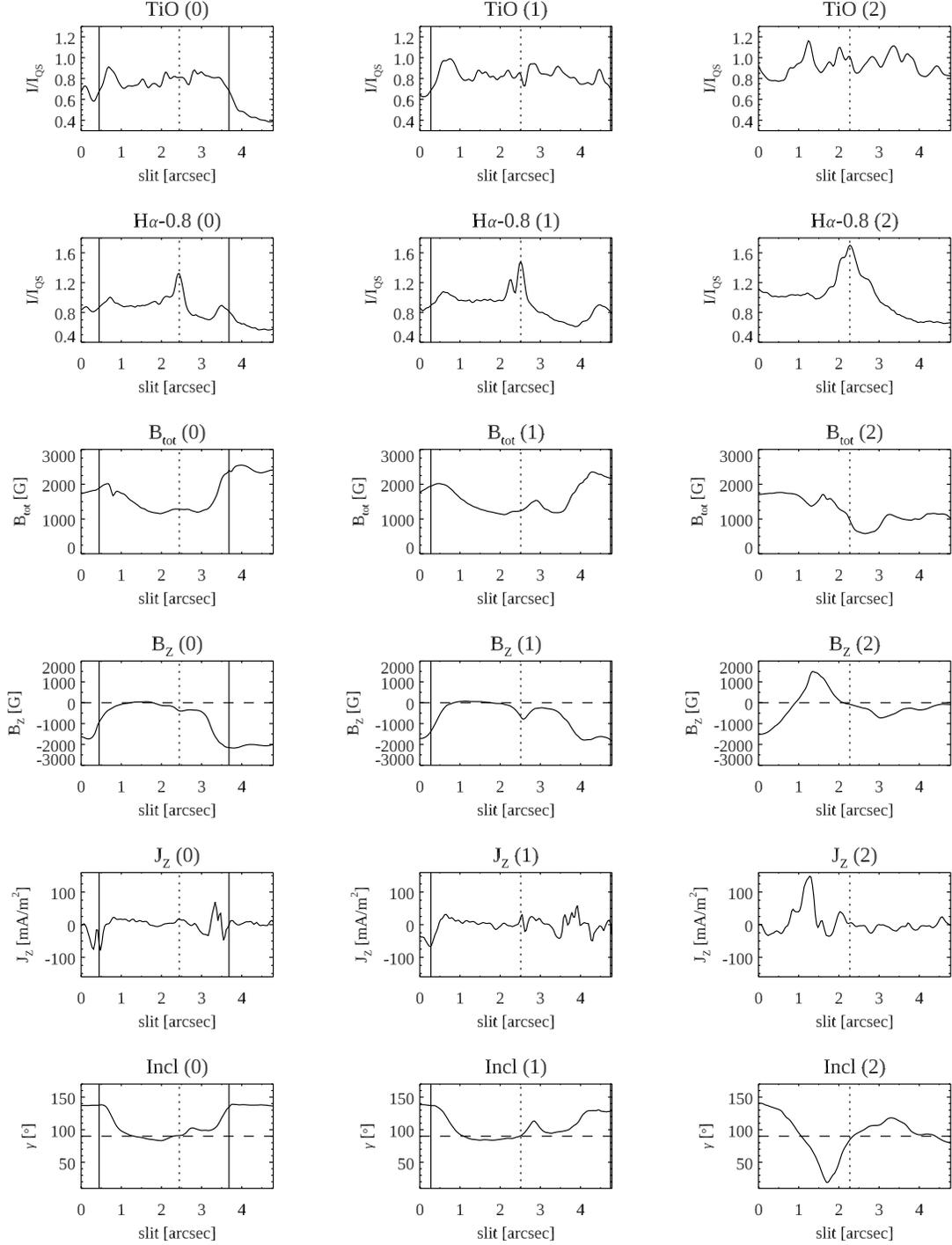}
    \caption{Profiles obtained along slits shown in the top-middle panel of Figure~\ref{tio_vis_niris}. The number inside each parenthesis indicates slit number. Solid vertical lines in profiles along slit 0 and 1 represents boundary of the LB defined as the $80\%$ level of the quiet region intensity. Dotted vertical line indicates the peak location of H$\alpha-0.8$\AA\, intensity profile.}\label{profiles}
\end{center}
\end{figure}

Figure~\ref{profiles} shows profiles of TiO and H$\alpha$ intensity, total and vertical magnetic fields, the vertical current density, and the inclination angles along several slits crossing the base brightening at different locations (indicated in Figure~\ref{tio_vis_niris} with numbers 0, 1, and 2). The slit locations were selected to cross the H$\alpha$ brightenings along the short axis of the LB, except for slit $2$, which crosses the most conspicuous and continuous brightening in H$\alpha$$-0.8$~\AA\,. The FSJ initially appeared at that location. {In Figure~\ref{profiles}, the slit number is indicated in parentheses for each data product.} The outer boundaries of the photospheric LB are well defined in the TiO intensity profile (solid vertical lines) along slits $0$ and $1$. The width of the LB is found to be around $3.\arcsec\,2$ ($3.\arcsec\,9$) in case of slit 0 (slit 1). The LB is brighter than the umbra but slightly darker than the quiet Sun region. We defined the edge of the LB to be at 69$\%$ level of the average quiet Sun intensity. It is well known \citep{Lit91, Lek97, Jur06} that the magnetic field of the LB is weaker than that of the ambient umbra (the total magnetic field strength in the LB is around $1300$~G in the case of panel $B_{\text{tot}}$(0)), and it is nearly horizontal (the inclination angle is between 80$\degr$ and 100$\degr$ in the panels Incl(0) and Incl(1)).

The dotted vertical line indicates the position of the peak H$\alpha-0.8$\AA\, intensity along the slit. In general, the intensity excess ranged from 30$\%$ (0) to 70$\%$ (2) of the quiet Sun intensity. As already mentioned above, in all three cases of slit positioning, the B$_z$ and the inclination profiles clearly show that the polarity inversion line was located in the vicinity of the H$\alpha$ brightening, while no significant increase of $J_z$ is observed at these locations. On the other hand, strong currents of 1.5~$\textrm{A~m}^{-2}$, which are higher than those at the LB boundaries, were detected at the slit 2 position x=1.\arcsec\,25, which is near the positive magnetic fields undergoing magnetic cancellation with the adjacent negative field of the sunspot (see the $B_z$ map in Figure~\ref{tio_vis_niris}). This finding indicates that the intense brightening and the FSJ may be caused by magnetic reconnection taking place in association with the LB's emerging flux.

The inclination angle of the umbral field in the vicinity of the LB is close to $140\degr$ and $130\degr$ in case of slits 0 and 1, respectively, which is already highly inclined. The electric currents are stronger at the location where the inclination angle rapidly decreases suggesting that the magnetic canopy may be located inside the apparent photospheric LB boundary. We also found that the fine strands of the FSJ originating from the western boundary of the LB are highly curved (Figure~\ref{vis_rd}). It appears that jet strands are directed along the canopy field rooted in the west umbral core, and the H$\alpha$ brightening is located just beneath the canopy field. We thus may estimate the lower limit of the height of the H$\alpha$ brightening to be $d\times \textrm{tan}\,40\degr$, simply assuming a constant inclination of the canopy field at $130\degr$ (inclination angle profile (1) in Figure~\ref{profiles}), and a distance $d$ between the projected H$\alpha$ brightening and the footpoint of the canopy field. Taking $d$=1.\arcsec\,5, we estimate the height of the H$\alpha$ brightening to be about 900~km, which corresponds to the lower chromosphere.

\section{SUMMARY AND DISCUSSIONS}\label{sec:discussion}

We performed analysis of vector magnetograms and imaging data obtained in the photosphere and chromosphere of the Sun using the 1.6~m GST aiming to explore the mechanism that produces fan-shaped jet-like ejections in LBs. More specifically, the temporal changes of the vector magnetic field and photospheric fine structures in an LB were investigated using NIRIS and TiO data, respectively. The dynamical properties of the FSJ and its base brightenings were explored using H$\alpha\pm0.8$\AA\, data from the VIS instruments. The following is a summary of our results.

\begin{enumerate}

\item We detected emergence of a compact positive polarity magnetic flux in a filamentary LB situated within a negative polarity sunspot. Magnetic cancellation of the emerging flux with the magnetic field of the sunspot proceeded at a rate of $5.7\times10^{18}~\textrm{Mx~hr}^{-1}$ ($4.4\times10^{18}~\textrm{Mx~hr}^{-1}$ measured for the sunspot flux). Vector magnetograms showed that the horizontal component of the emerged field was well-aligned with the LB orientation. Photospheric TiO images revealed that the filamentary structures in the emerging flux region, which are considered to be photospheric proxies of the horizontal field, are terminated along the boundary between the LB and the umbra. This indicates that the emerging field may not be a simple bipole but represents a complex magnetic structure directly connected to the umbra. The negative polarity footpoint of the emerging flux was less pronounced and spread over the LB boundary, whereas the positive flux is well concentrated at a single location.

\item Using the H$\alpha\pm0.8$\AA\, data, we detected jet-like ejections that last for about 20 minutes and recur every 15 to 20 minutes. The associated jet strands showed apparent horizontal drift along the longer axis of the LB at a speed {of \kms{2.0}} at the start of the ejection, then {\kms{18.4}} at the later phase. The observed FSJs were initiated at the vicinity of the newly emerging and canceling positive flux, and were accompanied by the elongated H$\alpha$ brightenings at the base.

\item The FSJs proceeded along the fine {filaments} that were rooted at the boundary between the LB and the west part of the umbra. The fine {filaments} were highly curved near the footpoints, suggesting that they may be directed along the umbral canopy field.
\end{enumerate}

One of the important results of this study is the detection of emergence and cancellation of the opposite magnetic polarity in a LB accompanied by chromospheric brightenings and recurrent plasma ejections. Many studies of LB jet-like ejections proposed magnetic reconnection as a preferable driver based on observed dynamic and accompanying brightenings. {Although these properties are consistent with a scenario of magnetic reconnection as a possible mechanism, firm evidence of the magnetic field change is still missing.} Based on the general magnetic field topology of a canopy structure in LBs \citep{Jur06, Rob16, Tia18} and MHD simulations of LB formation \citep{Tor15a}, it has been suggested that {the observed emergence and cancellation may be related to the emergence of an $\Omega$ loop emerging within LBs that reconnects with the overlying umbral fields.} However, very few direct observations of relevant magnetic field evolution were reported because of an insufficient spatial resolution or magnetic sensitivity. Only indirect evidences have been provided. For instance, \citet{Son17} reported a fine-scale photospheric pattern of diverging flows proceeding at a speed of about \kms{2} two minutes before the onset of plasma ejections, which was interpreted as evidence for magnetic-flux emergence leading to flux cancellation and ejection of chromospheric plasma. Here we present {for the first time} direct evidence for magnetic flux emergence and subsequent cancellation inside a LB, based on the high resolution. While magnetic cancellation itself may not be a sufficient evidence for magnetic reconnection, it indicates either an emerging U-loop or a submerging $\Omega$-loop through the photosphere. Yet, a combination of magnetic cancellation, chromospheric brightenings, and plasma ejections are widely accepted as an indicator of magnetic reconnection.

Secondly, we demonstrated the apparent horizontal broadening of the fan-shaped ejection, which has not been reported yet. The fan-shaped appearance of LB ejections is quite common, but its formation mechanism is not fully understood, even though it is crucial for the understanding the origin and the nature of these jets.
A recent study by \citet{Bai19} reported similar dynamics of fan-shaped ejections as well as propagation of a ``ribbon-like'' brightening at their footpoints observed in GST/VIS off-band H$\alpha$ data. Based on the morphological similarities between the observed ``ribbon-like'' brightening and standard flare ribbons, these authors suggested that the observed motion of the brightening may be a signature of slipping reconnection \citep{Aul06}. We favor their interpretation, because the observed extension of elongated H$\alpha$ brightenings can not be fully explained in terms of a single null point reconnection.

In that sense, the successive reconnection model is preferable for our observations. The question is why and where this successive reconnection takes place. We propose that slipping reconnection occurs within the quasi-separatrix layer \citep[QSL;][]{Pri95, Dem96, Dem97} that is formed between the LB emerging arcade and the overlying umbral canopy field. The well-pronounced magnetic cancellation between the positive flux P$_{\text{LB}}$ and the adjacent negative flux N suggests that the initial reconnection takes place between the (quasi-) open field anchored at N (the other footpoint may be far away from the negative polarity umbra) and the closed arcade field having one footpoint at P$_{\text{LB}}$. We recall that the emerging LB field was well aligned with the LB axis, indicating that the field lines may not be parallel to the overlying canopy field but they create some small angle. As the magnetic reconnection proceeds, newly reconnected field lines will drift upward toward the canopy field and straighten.

\begin{figure}[tb]
\begin{center}
    \includegraphics[width=0.95\textwidth]{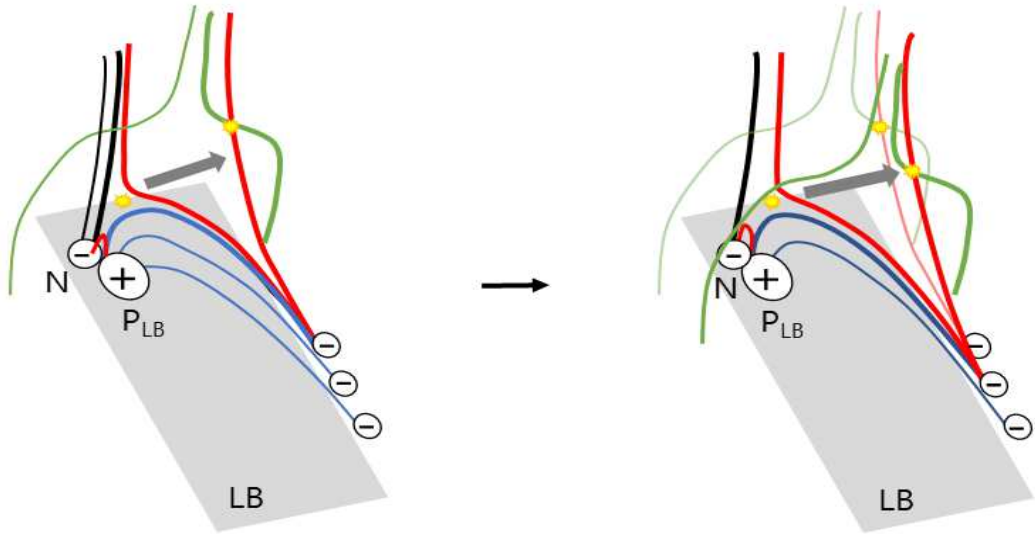}
    \caption{Illustration of suggested scenario of FSJ in the LB. Magnetic cancellation between P$_{\text{LB}}$ and N further will create new field lines (red lines) that will peel off the emerging dipole in the LB (blue concave-up loops) and move upward into the canopy area (green lines) above the LB. These field lines are likely to trigger there the slipping reconnection, where strong connectivity gradient will drive continuous exchange of the connectivity of neighboring field lines within the QSL. This process is then observed as apparent slippage of a field line itself and its footpoint along the complex QSL plane. Since this reconnection is fueled by the emerging field, many lines will `slip' along the canopy volume, so that one may observe continuous flow of field lines leading to expanding FSJ.}\label{cartoon}
\end{center}
\end{figure}

\citet{Aul06} pointed out that ``complex deformation of QSLs are typically expected when the boundary motions expand a quasi-connectivity domain in one region and shrink it in another region'' in 3D. Figure~\ref{cartoon} is a schematic representation of the proposed scenario for the formation of fan-shaped jet-like ejections in LBs. In our case, cancellation between P$_{\text{LB}}$ and N will create new field lines (red) that will peel off the emerging arcade (blue concave-up loops) and move upward into the canopy area (green lines) above the LB. These field lines, in turn, are likely to trigger slipping reconnection, where strong connectivity gradient, manifested as a local rotation of the magnetic field, will drive the continuous exchange of the connectivity of neighboring field lines within the QSL \citep{Pri95, Pri03, Aul06}. This process can be observed as apparent slippage of a field line itself and its footpoint along the complex QSL plane. Since this reconnection is driven by continuous flux emergence, many field lines will ``slip'' along the canopy volume, so that a continuous flow of field lines will be observed leading to the formation of expanding fan-shaped jets. We note that in Figure~\ref{cartoon}, the only part of canopy field lines are shown.

Assuming that the H$\alpha$ brightenings were formed via magnetic reconnection within the QSL, we estimated the lower limit of their height to be about 912~km based on the inclination angle of the adjacent umbral field and the projected transverse distance of the H$\alpha$ intensity maximum from the LB boundary. Based on this value, We speculate that the height of the QSL where the slipping reconnection took place to be about 900~km. \citet{Tia18} estimated the reconnection height to be 250--750~km based on the size of the inverted-Y shaped jets in an LB. Our estimation is in agreement with this report in a sense that reconnection most likely to occur in the lower chromosphere. The Anfv\'{e}n speed at height of $\sim 900$~km is estimated at about \kms{180} for a magnetic field strength of $500$~G and the hydrogen number density of $\sim 6\times10^{13}$~cm$^{-3}$. If we assume that the fan-shaped ejections result from slipping reconnection, then the speed of their horizontal drift along the LB may be comparable to the speed of the field line slippage. The measured drift speed was about {\kms{18.4}} which is slower compared to the estimated local Alfv\'{e}n speed. This may indicate that the reconnection process in our case is sub-Alfv\'{e}nic, or the reconnection height may be {a lot} lower than the estimate. More 3D simulations are required to understand the detailed process and the location of the slipping reconnection responsible for the LB fan-shaped ejection.

\acknowledgments
{The Authors appreciate the anonymous referee for his/her helpful comments on the manuscript.} This GST data set was also used in the study of \citet{Sah19}. This research was supported by basic research funding from Korea Astronomy and Space Science Institute (KASI). V.Y. acknowledges support from NSF AST-1614457, AFOSR FA9550-19-1-0040 and NASA 80NSSC17K0016, 80NSSC19K0257, and 80NSSC20K0025 grants. {JC and MM was supported by the Brain Pool Program of Korea (NRF-2019H1D3A2A0109943). The research of JC was supported by the Korea Astronomy and Space Science Institute under the R\&D program (Project No. 2020-1-850-07) supervised by the Ministry of Science and ICT.} BBSO operation is supported by NJIT and US NSF AGS-1821294 grant. GST operation is partly supported by the Korea Astronomy and Space Science Institute, the Seoul National University, and the Key Laboratory of Solar Activities of Chinese Academy of Sciences (CAS) and the Operation, Maintenance and Upgrading Fund of CAS for Astronomical Telescopes and Facility Instruments.
%
%
%

\newpage

\clearpage




\end{document}